# Determination of Boron Content in Tinkal Ore Samples with Geant4 using MCLLS Method

Onur ERBAY[a], Iskender Atilla REYHANCAN[a]

[a] *Istanbul Technical University, Energy Institute, Ayazaga Campus, Maslak, Sariyer, Istanbul*

Corresponding Author:

Onur Erbay

Istanbul Technical University, Energy Institute, Ayazaga Campus, Maslak, Sariyer, Istanbul, TURKIYE

Tel: +90 212 285 7393, email: erbayo@itu.edu.tr

## Abstract

A new Monte Carlo-Library Least Squares (MCLLS) method was developed for the neutron inelastic-scattering and thermal-capture analysis (NITA) technique in real-time online analysis systems to determine boron content in Tinkal ore samples. The Geant4 toolkit was used to simulate the Prompt Gamma-ray Neutron Activation Analysis (PGNAA) system for the NITA technique. Tinkal ore samples were bombarded with 14 MeV neutron beams in order to represent neutrons emitted from the Deuterium-Tritium (DT) neutron generator. After the simulation of shooting Tinkal ore samples with 14 MeV neutron beams was completed, histograms were obtained from the Geant4 simulation toolkit. Once the histograms are broadened according to the shape calibration of the BGO scintillation detector, the spectrum (broadened histograms) are ready to be used in the MCLLS process. A single element library was created by broadened histograms of the nine Tinkal ore samples. In addition, set #10 was selected to be an unknown sample and was used for validating MCLLS. After analysing the unknown sample spectrum, boron content was estimated by the single element library created by the MCLLS method. The estimated value of the boron content by MCLLS was consistent with the laboratory result.

**Keywords:** Monte Carlo Library Least Squares, Prompt gamma activation analysis, GEANT4, DT Neutron Generator

## 1. Introduction

Online elemental analyzers are used worldwide by mining and cement companies, as well as for academic and research purposes. Also, Prompt Gamma Neutron Activation Analysis (PGNAA) systems are assisting very important industrial processes. Besides that, PGNAA is a well-known, non-destructive and fast measurement method. The prompt gamma neutron activation analysis technique allows for the determination of light elements like H, B, C, O, and N. Moreover, F, Mg, Al, Si, Cu, Fe, P, and Zn elements are frequently determined by using 14 MeV neutrons (Révay et al., 2004).

In the prompt gamma neutron activation technique, the sample is bombarded continually by beams of neutrons (14 MeV neutrons). While some elements can be characterized by inelastic scattering reactions, thermalization of neutrons is needed for some elements. Such as, $^{10}B$ has a high cross-section for thermal neutrons at 477 keV of gamma-ray energy (Mendoza et al., 2018) and for that reason, a certain amount of sample thickness is needed in order to fast neutrons to be thermalized through the sample by inelastic scattering. Elements that go through inelastic scattering or neutron absorption reactions emit gamma-rays that are recorded by a gamma-ray spectrometer. Each element produces its own characteristic gamma-rays. These elements are identified by the energy of their gamma rays, and the intensities of the peaks at these energies indicate their concentrations in the sample. Gamma-ray spectrum are difficult to analyze, because they have non-linear parts and low resolution of gamma-ray spectrometer with BGO detector. Therefore, instead of the peak area method, the MCLLS approach was used in this study. W. Guo, R.P.Gardner and A.C. Todd., who introduced this method, express the advantages of this approach as follows; "Elemental library spectra are obtained by Monte Carlo simulation instead of experimental methods. Complex multiple peaks in the gamma-ray sample spectrum coming from different elements with close energy values are automatically resolved by this approach. The entire spectrum is utilized together. The measurement uncertainty is directly available from this approach" (Guo et al., 2004).

Geant4 (Agostinelli et al., 2003) which is used to obtain the single element library spectra of all the elements in the sample, is a multipurpose toolbox for simulating particle movement through materials. In addition, newly released G4NDL libraries show identical properties to the MCNP6 library. Geant4 which uses the Monte Carlo approach is suited for simulating experiments with updated libraries (Mendoza et al., 2018).

## 2. Experimental

PGNAA setup used in this study consists of three main parts. The first component is a Deuterium-Tritium (DT) neutron generator that generates 14 MeV neutrons, the second is a gamma-ray spectrometer with Ø3″×3″ BGO detector (Ortec DigiBase) and the final component is Tinkal ore samples. This setup (Fig. 1) was simulated by using Geant4. The histograms of the prompt gamma rays found as a result of the simulation were obtained by ROOT (Brun et al., 1996).

In Geant4, the ore sample is defined as box-shaped geometry as in the real experimental conditions. Simulation geometry of the experimental setup can be seen in Fig. 2 The rectangular shape at the bottom represents Tinkal ore. The contents of compounds in Tinkal ore samples were measured in Eskişehir, Kırka Eti Maden XRF laboratory. Results can be seen in 0. The amount of boron oxide in Tinkal ore samples used in this study varies from 23.1% to 30.8%. Humidity content of the samples were 8% and loss of ignition (LOI) value was 34%.

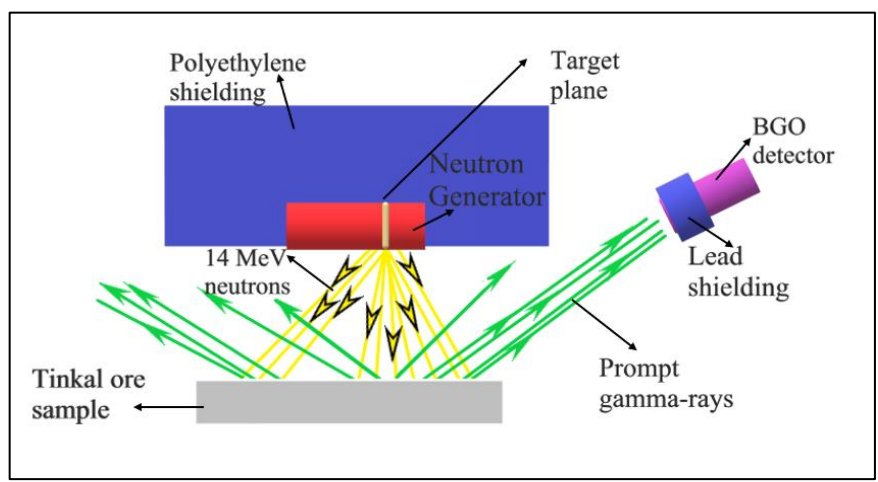

Fig. 1. A schematic representation of the real experimental setup.

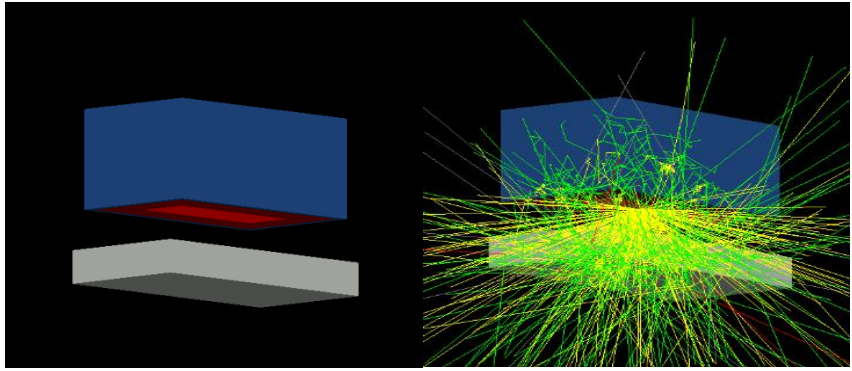

Fig. 2. Geant4 simulation of experimental setup.

Histograms of each Tinkal samples were obtained with Geant4. These histograms are needed to be broadened to match the response functions of the BGO detector. For this reason, energy and shape calibrations of the gamma ray spectrometer with BGO detector used in the experiment were conducted. Accurate measurements are crucial to most of the research. In this study, 551 keV, 661.7 keV, 1173.2 keV, 1332.5 keV, 2223.248 keV and 4438.94 keV gamma-ray peaks are used for energy (Fig. 3) and shape (Fig. 4) calibrations of BGO detector.

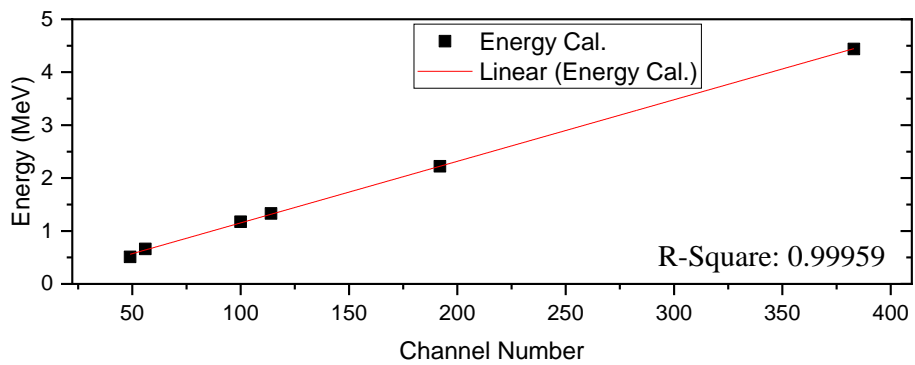

Fig. 3. Energy calibration of gamma-ray spectrometer.

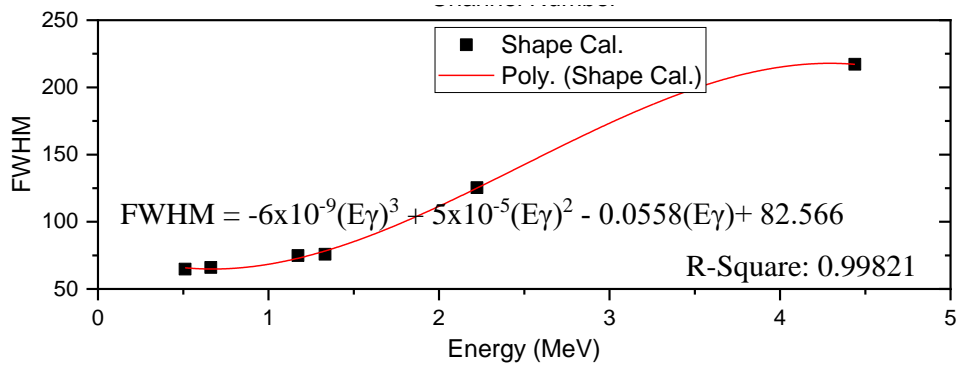

$$\text{FWHM} = -6\times10^{-9}(E\gamma)^3 + 5\times10^{-5}(E\gamma)^2 - 0.0558(E\gamma) + 82.566$$

Fig. 4. Shape calibration of gamma-ray spectrometer.

After completion of the simulation run for all sets, histograms are broadened according to the shape calibration parameters in order to construct a single element library. In Fig. 5, histogram of Set# 8 from Geant4 and spectrum (broadened histogram) can be seen.

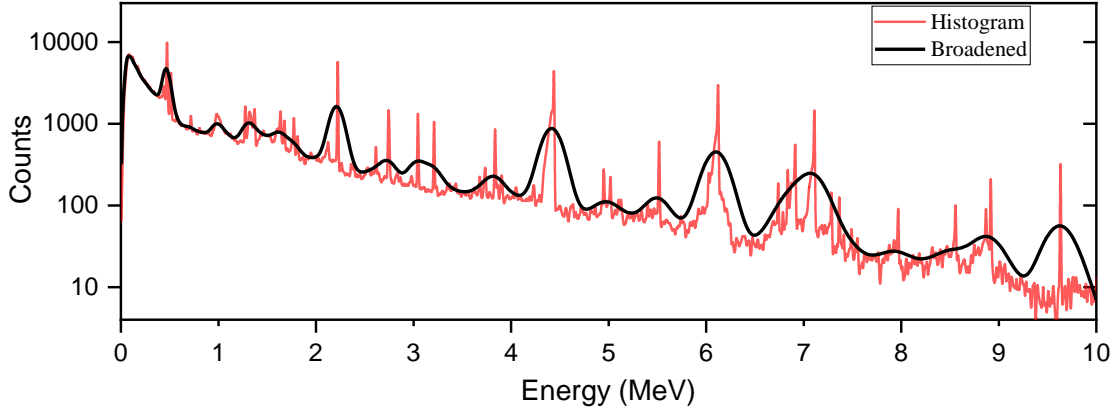

Fig. 5. Geant4 histogram and broadening.

Table 1. Content of Tinkal samples.(LOI value is 34%).

| Compound | Set1 (%) | Set2 (%) | Set3 (%) | Set4 (%) | Set5 (%) | Set6 (%) | Set7 (%) | Set8 (%) | Set9 (%) | Set10 (%) |
|---|---|---|---|---|---|---|---|---|---|---|
| $B_2O_3$ | 23.5 | 30.8 | 26.2 | 23.9 | 23.1 | 27.5 | 30.9 | 24.4 | 28.7 | 26.8 |
| MgO | 9.34 | 13.17 | 6.72 | 8.17 | 9.08 | 6.42 | 4.45 | 9.28 | 6.43 | 7.14 |
| $Na_2O$ | 9.18 | 5.15 | 11.35 | 9.87 | 10.02 | 11.38 | 13.59 | 10.79 | 11.62 | 11.55 |
| CaO | 7.79 | 4.13 | 7.97 | 9.22 | 8.13 | 6.75 | 4.33 | 11.2 | 6.30 | 6.77 |
| $SiO_2$ | 7.62 | 3.17 | 5.25 | 6.22 | 7.13 | 5.08 | 3.48 | 7.54 | 6.29 | 5.23 |
| $Al_2O_3$ | 0.56 | 0.13 | 0.23 | 0.25 | 0.40 | 0.22 | 0.20 | 0.33 | 0.50 | 0.24 |
| SrO | 0.43 | 0.25 | 0.38 | 0.39 | 0.42 | 0.30 | 0.24 | 0.50 | 0.30 | 0.35 |
| $SO_3$ | 0.16 | 0.06 | 0.14 | 0.18 | 0.18 | 0.12 | 0.09 | 0.24 | 0.14 | 0.15 |
| $Fe_2O_3$ | 0.12 | 0.03 | 0.054 | 0.06 | 0.09 | 0.05 | 0.05 | 0.09 | 0.10 | 0.05 |

## 3. MCLLS Method

This mathematical method is based on the principle of evaluation of the unknown sample spectrum which is taken as the sum of the library spectrum of each element. The least-squares criterion is used to find the most probable values of relative elemental contents in the simulated Tinkal samples. The sum of the library spectrum of a single element is represented in Eq. (1), where Eq. (2) represents the least squares norm to define the most apparent values of related elemental contents in the measured Tinkal samples.

The fundamental supposition of the MCLLS method is that the spectra of an unknown sample can be divided into parts that belong to the monoenergetic segments. Based on this approach, the single element library constructed by LLS can be used to perform the reverse matrix operation (Wang et al., 2012) to determine compound contents in the spectra of an unknown sample.

$$y_i = \sum_{k=1}^{9} a_{ik} x_k + e_i, \quad i = 1,2,3,\dots,1024 \quad (1)$$

$$M = \sum_{i=1}^{1024} \omega_i \left( y_i - \sum_{k=1}^{9} a_{ik} x_k \right)^2 \qquad (2)$$

In Eq. (2) $\omega_i$ is the statistical weight for channel $i$, $y_i$ is the count rate in channel $i$ for the unknown sample spectra, $a_{ik}$ is the count rate in channel $i$ for the single-element spectrum of element $k$, and $x_k$ is the relative content of element $k$. It is expected that the $a_{ik}$ does not change with the elemental contents, within the MCLLS approach. As a result, a partial derivative with respect to some relative content $x_k$ can be used to find the minimum of the parameter $M$. The derivative of $M$ is then set as zero (Eq. 3):

$$\partial M / \partial x_k = \sum_{i=1}^{1024} \omega_i \left( y_i - \sum_{k=1}^{9} a_{ik} x_k \right) a_{ik} = 0 \qquad (3)$$

Eq. (3) can be transformed into matrix model:

$$A^T \omega y - (A^T \omega A) X = 0 \qquad (4)$$

and resolving Eq. (4) that for X yields:

$$X = (A^T \omega A)^{-1} A^T \omega Y \qquad (5)$$

Element contents can be find with Eq. (5).

## 4. Results and Discussion

In this study, $Al_2O_3$, $SrO$, $SO_3$ and $Fe_2O_3$ were not performed because their contents were less than 3%. According to the $B_2O_3$, $MgO$, $Na_2O$, $CaO$ and $SiO_2$ contents of unknown Set# 10, laboratory and MCLLS results were compared in 0. A single element library was created by using nine Tinkal samples spectrum with the MCLLS method. As seen from Table 2, nine samples are enough to determining boron oxide content with good accuracy. Content values of other compounds were given to demonstrate the abilities of the MCLLS method and Geant4.

Table 2. Comparison of the MCLLS and Laboratory results.

| Compound | MCLLS Results | Laboratory Results | Difference (ABS) | Percentage % |
|---|---|---|---|---|
| $B_2O_3$ | 27,55 | 26,80 | 0,75 | 2,77 |
| MgO | 7,45 | 7,14 | 0,31 | 4,27 |
| $Na_2O$ | 10,80 | 11,55 | 0,75 | 6,74 |
| CaO | 6,32 | 6,77 | 0,45 | 6,91 |
| $SiO_2$ | 5,70 | 5,23 | 0,47 | 8,55 |

In any PGNAA system, the spectrum obtained from gamma-ray spectrometer has components of pulse pile-up distortions and sum peaks. These effects cause non-linear problems in spectrum analysis and were not studied in this article.

## 5. Conclusion

In this study, the MCLLS method was developed for PGNAA systems to determine boron content in Tinkal ore samples. It is found that a single element library created by the MCLLS method was successful in determining the boron content of unknown sample. Also, the results of the MCLLS method have good accuracy and are consistent with laboratory results. In simulation, non-linearity problems are not considered. However, those problems can be minimized by the iteration process till single element library values comply with the content of the broadened histograms. For real-time online characterization of bulk material, the MCLLS method has the property of being applicable.